# Gender differences in research areas, methods and topics: Can people and thing orientations explain the results?[1]

Mike Thelwall, Carol Bailey, Catherine Tobin, [School of Mathematics and Computer Science, International Academy, School of Sciences], University of Wolverhampton;
Noel-Ann Bradshaw, Department of Mathematical Sciences, University of Greenwich.

Although the gender gap in academia has narrowed, females are underrepresented within some fields in the USA. Prior research suggests that the imbalances between science, technology, engineering and mathematics fields may be partly due to greater male interest in things and greater female interest in people, or to off-putting masculine cultures in some disciplines. To seek more detailed insights across all subjects, this article compares practising US male and female researchers between and within 285 narrow Scopus fields inside 26 broad fields from their first-authored articles published in 2017. The comparison is based on publishing fields and the words used in article titles, abstracts, and keywords. The results cannot be fully explained by the people/thing dimensions. Exceptions include greater female interest in veterinary science and cell biology and greater male interest in abstraction, patients, and power/control fields, such as politics and law. These may be due to other factors, such as the ability of a career to provide status or social impact or the availability of alternative careers. As a possible side effect of the partial people/thing relationship, females are more likely to use exploratory and qualitative methods and males are more likely to use quantitative methods. The results suggest that the necessary steps of eliminating explicit and implicit gender bias in academia are insufficient and might be complemented by measures to make fields more attractive to minority genders.

## 1   Introduction

Gender differences are prevalent in academia. Females study, lecture and research some fields more than males but are heavily underrepresented in most mathematically intensive areas (e.g., Holman, Stuart-Fox, & Hauser, 2018). The reasons for this imbalance are unclear. Explicit bias against women and any gender differences in mathematical performance cannot explain much of the continuing inequality (Ceci & Williams, 2011; Justman & Méndez, 2018). Instead, differing career choices stemming from female people-related interests and male thing-related interests is a plausible partial explanation (Su, Rounds, & Armstrong, 2009; see also: Lordan & Pischke, 2016; Lippa, 1998; Lippa, Preston, & Penner, 2014). Thing and people dimensions do not fully account for the low numbers of women in engineering, mechanics and computing, or the high numbers of women in medicine (Su & Rounds, 2015). They also do not explain why women are over-represented in life sciences, which do not necessarily involve people, but under-represented in dentistry and surgery, which are people-based. Computer science, engineering and physics may be avoided by females because their male dominated cultures are (inadvertently) unattractive to women, they have "chilly climates" that alienate females (for other subjects, see: Britton, 2017; Stockard, Greene, Richmond, & Lewis, 2018), there is implicit bias within primary and secondary education (Robnet, 2016), girls lack necessary early exposure to the topics and females have lower self-efficacy for them (Cheryan, Ziegler, Montoya, & Jiang, 2017). Disciplinary cultures may therefore partly explain field gender compositions. Differing abilities of fields to satisfy personal goals for status or social impact is another likely influence (Yang & Barth, 2015). Nevertheless, a deeper understanding of gendered factors influencing academic career choices is needed if gender imbalances are to be redressed. It is also important to understand why females have successfully overturned male domination in some disciplines. A fine-grained examination of gender differences in research specialisms and topics within academia is a pre-requisite for any deeper investigation.

As reviewed below, previous research into the relationships between interests, gender and academic careers has relied on survey data from children and adults. It has several limitations that need complementary approaches to address.

a. Prior studies have mostly relied upon survey data using interest categories that do not match the wide range of academic fields, each of which has its own peculiarities.

b. It is not known whether there is a fundamental neurological or social psychological explanation for the people or things dimensions so that they could be aligned with more specific factors. More detailed evidence from the variety of academic specialisms may help with this.

c. Matching interests declared in surveys (e.g., things/people) to academic discipline gender makeup (Su, Rounds, & Armstrong, 2009) may obscure characteristics of female researchers that manifest themselves within fields rather than through field choice.

The current study uses a novel approach to assess whether academic publishing is consistent with current theories of gender differences. Although gender differences in academic publishing is an important issue (Aksnes, Rorstad, Piro, & Sivertsen, 2011; Ceci & Williams, 2011; Larivière, Ni, Gingras, Cronin, & Sugimoto, 2013), the primary goal is to gain insights into underlying career choices. This paper analyses the publications produced by working academics and uses the words in these publications as a proxy for their interests. It therefore takes a different perspective from the most insightful prior studies (Ceci & Williams, 2011; Cheryan, Ziegler, Montoya & Jiang, 2017; Su, Rounds, & Armstrong, 2009; Su & Rounds, 2015; Yang & Barth, 2015) by examining what researchers write and ignoring the social, biological and economic factors that shape career choices. Focusing on researchers' outputs allows a data-driven analysis strategy that circumvents the constraints of surveys, allowing narrow academic fields and topics to be analysed. This approach cannot directly address cause-and-effect relationships, however, and cannot draw firm conclusions about underlying career choices. The goal is to analyse gender differences in academic publishing and look for *insights* into topic and method-based factors affecting underlying career choices, by describing current gender differences in topics and methods within scholarly publishing and assessing the extent that existing theories (e.g., people/things) can account for them. Thus, gender differences found in the data (research fields or topics) that are not consistent with prior theories would point to potential problems with these theories as explanations for career outcomes. Gender differences are discovered with a word frequency driven exploration of factors that associate with researcher gender across multiple narrow fields in the USA. The word frequency approach produces lists of terms that associate with one gender in articles from 2017 in many of the 285 Scopus narrow fields examined. These gendered terms are then investigated for agreement with previously hypothesised gender differences. Whilst text analysis has been used to deliver insights into individual behaviours (Tausczik & Pennebaker, 2010) and within individual fields (Elsevier, 2015; Nielsen, Alegria, Börjeson, et al., 2017; Vogel & Jurafsky, 2012), none have been peer-reviewed journal articles and this is apparently the first attempt to use it to gain insights into gender differences across academic publishing.

## 2 Literature review

The people and thing dimensions that have been proposed as explanations for academic career choices have also been investigated in the wider context of general career choices. The current gender imbalances in science may be partly due to gendered career choices by those that have selected an academic field. It is therefore important to understand how gender affects career choices in general.

### 2.1 Gendered career choices

In 2016 in the USA, women comprised a minority (46.8%) of the labour force but a higher proportion (15.9% compared to 13.7% for males) of females had an advanced degree (above bachelor's level) (DOL, 2018a). Fewer women with an advanced degree were employed (72.1%) than similar males (77.6%) (DOL, 2018b). Parenting does not affect the employment rate of males in the USA but reduces it for females, particularly with a child under 5 (DOL, 2018b). This choice may be socially constrained and may be influenced by males not wishing to take career breaks for childcare, being in a higher paid job than a female partner, or considering childcare breaks to be gender inappropriate. Female career breaks may have the longer-term effect of turning fathers into the main earners in households, especially if mothers are not supported well when returning to work. Females may spend fewer years in the workforce due to taking more time off for carer roles, or taking early retirement alongside an older male partner (although women are starting to work longer: Goldin & Katz, 2018). Against this background of shorter effective working lives for females, gender equity for any given career may correspond to females constituting slightly less than 50% of employees unless the wider social issues are addressed by the employer or society. For example, if females have a working life of 1 year less than males due to retirement age decisions and career breaks and part-time working to cover for carer responsibilities, they would constitute 49% of the national workforce. The average age of working males (50.5) in the USA is five years older than the average age of working females (45.5) in one not fully representative data set (combining the paid employment and self-employed categories in: Federal Reserve, 2018).

In contrast to the minor gender differences in abilities (Ceci & Williams, 2011), there are large gender differences in employment types in the USA, with the most common jobs for women in 2015 being teachers, nurses, secretaries/administrators, customer service, managers, retail supervisors, cashiers, office supervisors and accountants/auditors. In contrast, the most common male jobs are drivers, managers, retail supervisors, labourers, retail sales, janitors and cleaners, software developers, sales representatives, and groundskeepers. The most gender biased jobs include secretaries/administrators (94.5% female), nursing (89.4% female), preschool teachers (96.8% female), elementary/middle school teachers (80.7% female), drivers (5.1% female), labourers (2.9% female), carpenters (1.8% female), car mechanics (1.5% female), and electricians (2.3% female) (DOL, 2018c).

Whilst part of the reasons for gender differences in employment may be employer gender bias, it seems likely that they are driven by choices that are both socialised and made within social constraints and pressures to conform to gender identities (Dinella, Fulcher, & Weisgram, 2014). As an example of a social constraint, teaching may be attractive for females with young families seeking time off in the school holidays to look after their children. The higher rates of part time working amongst females may be partly due to a greater need to fit around unpaid carer roles (FCA, 2016).

Why do females have substantially different career choices to males overall? Following cognitive social learning theory, numerous small rewards and punishments during childhood and into adulthood shape interests along gender normative dimensions (Bussey & Bandura, 1999; Hyde, 2014). People may also internalise gender differences that they perceive in their environment, presume them to be powerful and use them to constrain or guide their choices in directions that appear to be beneficial for them (Wood & Eagly, 2012). This may lead people to pursue career goals that fit within their internalised beliefs about what would be regarded as positive for their gender role. In terms of people/thing orientations, one or both may even have a biological component (Beltz, Swanson, & Berenbaum, 2011; Hines, 2011). Thing and people gender biases have also been previously discovered in written texts (object properties vs. psychological and social processes: Newman, Groom, Handelman, & Pennebaker, 2008).

## 2.2 Gender differences in academic-related interests

Females in the USA have outnumbered males in college enrolments since the 1980s (Goldin, Katz, & Kuziemko, 2006) but are substantially underrepresented in engineering, mathematics, computer science, and physics (EMCP) (Hyde, 2014). At the PhD level, in 2013-14 in the USA, women accounted for 21% in computing (NCES, 2018a), 23% in engineering (NCES, 2018b), 33% in physical sciences, and 29% in mathematics and statistics (NCES, 2018c). Women are close to parity or ahead in other STEM areas, including 53% within the biological and biomedical sciences (NCES, 2018d). Although women are under-represented in mathematics-oriented scientific fields (geosciences, engineering, economics, mathematics, computer science, physical sciences) and over-represented or at parity in the life sciences, psychology and social sciences, this cannot be explained by differences in mathematical ability (Ceci, Ginther, Kahn, & Williams, 2014). Males and females have similar mathematical capabilities (Lindberg, Hyde, Petersen, & Linn, 2010) but there is some evidence of males outperforming females at the extreme of mathematics performance (Ceci, Williams, & Barnett, 2009) and on average maths skills when leaving school (ACT, 2014), whilst girls achieve a higher Grade Point Average at high school in mathematics (ACT, 2014). These differences (diminishing over time) are not enough to account for the extent of the differences in career choices. Bias against women also does not seem to be a major cause of the current differences in most areas (Ceci, Ginther, Kahn, & Williams, 2014). The greater need for flexibility at work (Goldin, 2015) may also impact on the types of fields that retain females, as gender differences in the perceived value of academic fields (Eccles, 2007), and perhaps also differences in general capabilities relevant so some fields, such as visual skills (e.g., Baker & Cornelson, 2018).

In 2007, the America Competes Act legislated for increased investment in STEM research and education for students of all years. An amendment three years later aimed to increase the number of underrepresented minorities in STEM fields (Blackburn, 2017). This led to initiatives such as Obama's Educate to Innovate campaign in 2009. Similar initiatives targeting young female students may result in more girls opting for STEM-related majors. For later success in STEM, it is important for the student to have identified with it at a young age (Bieri Buschor et al., 2014; McCarthy & Berger 2008; Valla & Williams, 2012) but, if these initiatives work, it may take time to change male-dominated cultures in higher education STEM departments (Beede et al, 2011).

Males are underrepresented in undergraduate education for health care, elementary education and the domestic sphere (HEED) subjects, with the main reason hypothesised so far being cultural rather than self-efficacy: Males were more likely to feel that they did not belong in HEED subjects (Tellhed, Bäckström, & Björklund, 2017). The lower overall academic performance of males at school (Schoon & Eccles, 2014) may also constrain their career choices.

Gender differences in interests are more substantial than differences in abilities or psychological characteristics (Su, Rounds, & Armstrong, 2009). Masculine cultures (Cheryan, Ziegler, Montoya, & Jiang, 2017) and differences in interest are arguably the main current cause of gender differences in career choices within academia in the USA, rather than prejudice or explicit bias or any differences in abilities (Ceci & Williams, 2011; Su, Rounds, & Armstrong, 2009). These choices continue within careers (the leaky pipeline hypothesis: Clark Blickenstaff, 2005), increasing the gender imbalance in mathematics-related areas. Industrialised nations emphasise the importance of creative fulfilment within a job, which may transfer gender differences in interest into gender differences in careers (Charles & Bradley, 2009). The low participation rates of women in mathematics-intensive fields may therefore reflect a lack of interest in them. Mathematics-intensive fields might also be seen as masculine areas, socially penalising participating females (Wood & Eagly, 2012) even though practising researchers may have strategies to mitigate this (Richman, Vandellen, & Wood, 2011).

Su, Rounds, and Armstrong (2009) found a tendency for males to prefer working with things and for females to prefer working with people in their meta-analysis. They found a weaker tendency for females to be interested in social and artistic activities and males to be substantially more interested in engineering and moderately more interested in maths and science. Female interests in people may compete with their interest in things for career choices. In other words, given a male and a female with the same level of interest in things, the female is more likely to have a higher interest in people and opt for a more people-related career (Su, Rounds, & Armstrong, 2009). Thus, the lack of women in thing-related careers may give an exaggerated impression of their (average) lack of interest in things. The people and things dimensions are sometimes treated as opposite ends of a spectrum but they are two separate dimensions with little or no correlation (Graziano, Habashi, Evangelou, & Ngambeki, 2012; Graziano, Habashi, & Woodcock, 2011; Woodcock, Graziano, Branch, Habashi, Ngambeki, & Evangelou, 2013).

An attempt to explain gender differences in STEM subjects using role congruity theory argues that the ability of a job to fulfil a person's goals is important. Males and females may differ in the extent to which they have *communal* (e.g., positive social impact, compatibility with family life) or *agentic* (e.g., enhancing personal status) goals and may choose jobs that they believe match their goals, or leave jobs that do not match them (Diekman, Steinberg, Brown, Belanger, & Clark, 2017; Diekman, & Steinberg, 2013; Diekman, Brown, Johnston, & Clark, 2010). Similarly, power has been previously noted as a male-associated career goal (Gino, Wilmuth, & Brooks, 2015) and so males may be more interested in subjects such as law, management and organisation. There is an overlap between the communal goals theory and the people and things theory since agentic goals are perceived to be more likely to be satisfied by thing-related jobs, whereas communal goals are perceived to be more likely to be satisfied by people-related jobs (Yang & Barth, 2015). There is some evidence that the two models do not fully overlap, however, with the perceived ability of a job to support status and social impact goals improving the prediction of job preferences compared to people/thing interests alone (Yang & Barth, 2015).

A fine-grained analysis of the relationship between gender, interests and Science Technology Engineering and Mathematics fields (excluding the arts and humanities, but including social and health sciences) investigated 13 broad areas, five of which are only partly academic-oriented (Su & Rounds, 2015). Abilities and mathematical content were not found to influence the gender composition of these 13 broad areas and the best two predictors of gender composition were the degrees of people orientation and thing orientation in each area. The low female numbers in quantitative fields was hypothesised to be due to these fields' greater focus on things and lesser focus on people. The exclusion of some academic fields is a limitation of this study, as is the use of a standard occupational categorisation scheme that has relatively old roots (The Holland Codes (Spokane, Luchetta, & Richwine, 2002), with origins in the 1950s before the information technology revolution), was not designed to address people and thing dimensions (the Realistic and Social categories were adopted for this) and was not designed for academia.

The degree of thing/people orientations is insufficient to explain the extent of absence of females from Computer Science, Engineering and Mechanics areas, nor the extent of the dominance in Medical Services (Su & Rounds, 2015). The situation is complicated by people and thing orientations varying within broad fields. For example, the male dominated area of Computer Science includes people-related academic specialities, such as computer-human interaction and natural language processing, even if they do not necessarily involve face-to-face interaction.

The low rates of female participation in computer science, and engineering could be due to their masculine cultures (Cheryan, Ziegler, Montoya, & Jiang, 2017) but it is not clear why such cultures have persisted in these but not others, given that almost all areas of academia were originally male dominated or exclusively male. Computer science, engineering and physics seem to have male-associated cultures that females may dislike or not want to be associated with (Cheryan, Ziegler, Montoya, & Jiang, 2017). Lower female self-efficacy for these fields, irrespective of skill level, and a lack of early exposure may also reduce the number of females participating in them (Cheryan, Ziegler, Montoya, & Jiang, 2017). Thus, cultural, psychological and educational factors probably influence gendered academic discipline choices in a way that is not necessarily related to gendered levels of interest in the object of study.

A prior text-based analyses of gender differences in research specialisms for 11,931 international authors within computational linguistics found a greater female focus on dialog, discourse, and sentiment in contrast to parsing, formal semantics, and finite state models. This workshop paper used a version of topic modeling for an arbitrarily selected number of topics (100, including 27 discarded as "unsubstantive") and subject experts manually labelled each topic based on its ten most closely associated words (Vogel & Jurafsky, 2012). Unpublished topic modelling studies have also investigated management science and sex/gender topic differences within medical research (Nielsen, Alegria, Börjeson, et al., 2017). An (unrefereed) analysis of the gender composition of German research teams (Biochemistry, Genetics and Molecular Biology: 48% female; Physics and Astronomy: 19% female) 2010-2014 compared "key phrases" (not precisely defined, but extracted with the "Elsevier Fingerprint Engine" using "a variety of Natural Language Processing techniques to the titles and abstracts") from male-only and female-majority authored publications, comparing them with a visual inspection of a co-occurrence network diagram. Suggested differences included a male association with "probability" and "theoretical models" in contrast to a female association with "family", "child", "women", "infant", and "pregnancy" (Elsevier, 2015). Although not statistically robust (Leydesdorff & Nerghes, 2017) these methods have suggested that there are gender differences in topic choices within broad fields. Another unrefereed analysis (and closest in scope to the current paper) used variety of bibliometric techniques to investigate gender (based on first names in Scopus, when present) in research publishing across academia 1996-2015 in 11 countries and the EU (28 countries combined). It reported the proportion of male and female authors 1996-2000 and 2011-2015 in each of 27 broad subject areas, showing an increasing proportion of females in the second period (Figure 1.3 of: Elsevier, 2017). This paper also reported a topic analysis of the gender research topic (Chapter 3 of: Elsevier, 2017).

## 3   Research questions

This study seeks factors that associate with female participation in research (not just STEM) to clarify and expand on the previously hypothesised thing and people interest dimensions. It ignores all forms of participation in academia that do not lead to Scopus-indexed journal articles. This issue is addressed from three perspectives: choice of field to research; choice of topic to research and research methods; choice of topic or methods relative to the chosen field. Whilst the first of these is the standard approach, the other two address the issue from different perspectives and may give cross-field insights into female choices and working practices. The underlying hypothesis is that researchers tend to have some control over what to research (both field and topic), even if there are constraints. The research questions are as follows.

1. What are the current (2017) gender difference in US participation rates for **broad and narrow academic fields**?
2. Which research topics and methods have gender differences in uptake by US researchers (**across academia)**?
3. Which research topics and methods have gender differences in use by US researchers **within narrow academic fields**?
4. Are the results consistent with the people/thing theory, do they suggest refinements and do they point to important exceptions?

The second and third research questions are different because a topic or method could be gendered across academia by females tending to choose fields for a research topic or method (e.g., nursing), whereas a topic or method could be gendered within multiple academic fields by females being more likely to choose a research style, method or theme (e.g., qualitative) within their chosen field.

## 4   Methods

The research uses a mixed methods approach, following quantitative methods as far as possible and then switching to qualitative analyses when no quantitative methods will give useful results. This is in sharp contrast to much applied psychology research into factors affecting occupational preferences that is purely quantitative (e.g., ANOVA, regression, confirmatory factor analysis), with the main qualitative contribution being an interpretation of variables or factors (e.g., that a factor represents a "thing orientation"). Here, the switch to a qualitative approach is necessary for a meaningful analysis of the data, albeit at a loss of generalisability and rigour. Also in contrast to similar psychology research, the data used is from career outputs (journal articles) rather than questionnaires and so self-reported gender data is not available, and there is little chance to detect cause-and-effect relationships. The data here is much larger than in typical psychology research, however, and is analysed at a greater level of detail (285 narrow fields).

The research design incorporates a common data set that is gathered and analysed in the following six steps that are first summarised and then described in detail.

1. *Multidisciplinary gendered US journal article datasets creation*: Generates a large set of research articles from a comprehensive set of narrow academic fields with US first author genders as the raw data. The fields go beyond STEM subjects to include the arts and humanities. A second version of this dataset is created after removing duplicate articles appearing in multiple fields (RQ1,2,3).
2. *Broad and narrow field first author gender summarisation*: Reports the proportion of male and female authors in broad and narrow fields (RQ1).
3. *Gendered academic terms extraction*: Uses word frequency analyses to identify terms that associated with either males or females across academia (RQ2).
4. *Field-specific gendered terms extraction*: Uses word frequency analyses to identify terms that associated with either males or females within each individual field (RQ3).
5. *Common academic gendered terms extraction*: Identifies terms that are gendered for multiple fields (RQ3).
6. *Gendered academic theme discovery*: Investigates common gendered terms for underlying themes (RQ2,3).

Articles were assumed to be mainly based on the research conducted by the first author. This is true in most broad fields (Larivière et al., 2016), but may be less true in Economics, Social Sciences and Maths due to sometimes using alphabetical authorship order (Levitt & Thelwall, 2013). The estimated number of gender errors due to alphabetisation is 2% overall, with an estimated maximum of 14% for any narrow field with over 50 articles (Accounting). The net impact of this on the first results table calculations for the worst case, Accounting, on field participation rates is only 1.1%, however. This small net impact occurs since about half (54%) of the Accounting gender swaps are female-to-male, which largely cancel with the slightly fewer male-to-female errors. The gender errors reduce the power of the word association tests described below (relevant to the remaining results tables) but do not invalidate them. Last authors may be more senior in biomedical research (Holman et al., 2018) and may be influential in the choice of topic to research but it is still relevant to analyse the first author alone because they are likely to have contributed most to the study and have at least agreed to participate in it, even if they did not choose it. Moreover, a survey of multi-authored medical article authors found the first author to be most likely to fulfil accepted authorship criteria, with the last author being seven times less likely have done so (Zbar & Frank, 2011).

## 4.1 Step 1: Multidisciplinary gendered US journal article dataset

Up to 10,000 records for documents of type journal article with a United States author country affiliation in all 310 Scopus narrow fields published in 2017 were downloaded from Scopus in January 2018, a total of 508,283 articles. This is a complete set of Scopus narrow fields, from Agricultural and Biological Sciences (miscellaneous) to Speech and Hearing, excluding the multidisciplinary set since this does not have a topic focus. Each narrow field falls within one of 26 broad fields, covering the full spectrum of academic research. Narrow fields are classified with the Scopus All Science Journal Classification (ASJC) codes (http://ebrp.elsevier.com/pdf/Scopus_Custom_Data_Documentation_v4.pdf).

The Scopus classification scheme is applied manually to journals, placing them in one or more narrow subject categories based on the typical content of articles. This categorisation is conducted by an interdisciplinary group of subject experts organised by Elsevier. Although it seems to be primarily designed for information retrieval, it is also used by Elsevier for impact calculations for journals and in its commissioned reports (e.g., Elsevier, 2013). Whilst the classification scheme rarely omits journals from relevant categories, it sometimes also places journals into categories in which they have little relevance (Wang & Waltman, 2016). For the analysis here, there is a small risk that a large journal from a Scopus narrow field influences the results of another narrow field with a different gender perspective.

The restriction to a maximum of 10000 journal articles per field was due to a limit of 5000 per query in Scopus, which allows records to be downloaded in chronological and reverse chronological order. For the 10 fields with over 10,000 articles, the set represents the first and last 5000 articles indexed by Scopus within this field. This is a time-balanced set but it is possible that the missing records in the 10 partial fields are not representative of the gender breakdown of the retrieved records. It seems unlikely that this bias would be systematic enough to influence the overall results, although it might influence the outcomes for individual fields.

Articles with a first author not from the USA were removed. Gender was detected using a set of 1021 male and 3937 female names, each of which is used 90% of the time by one gender in the 1990 US census (e.g., see: Larivière, Ni, Gingras, Cronin, & Sugimoto, 2013). Records with an unknown first author gender (43%) were removed. This high proportion is partly due to the necessity to remove relatively gender ambiguous or unisex names as well as rare names (for which reliable gender statistics are not available) in order to have a low proportion of false matches. The use of first names to approximate gender may slightly over represent males since a greater proportion of females have gender ambiguous first names (Lieberson, Dumais, & Baumann, 2000). This method has the advantage of transparency in contrast to online gender identification API services, and so it can be reproduced and its limitations understood. Gender APIs typically rely on names in social web sites that may be informal (e.g., Pat for Patrick or Patricia; Ali for Alison, Alistair or Ali), increasing the error rate. Nevertheless, the US census method used here biases the results against academics from cultures (e.g., Sikh) using gender neutral first names, or that were not well represented in the USA in 1990. A prior academic gender analysis paper used a variety of other sources to identify name genders because they analysed articles from multiple countries (Larivière, Ni, Gingras, Cronin, & Sugimoto, 2013). It is not possible to use non-US sources to investigate US names because of international variations in the gender orientation of names (e.g., Honey is a male name in India and Andrea is a male name in Italy) that would add false matches, despite increasing the proportion of authors classified.

Some (8%) of the first authors in the dataset could not be classified for gender because they used initials rather than names. An additional 8% of the authors could have been classified using gender API information relying on social media profiles (using a rule of at least 90% monogender based on at least 100 social media profiles) but the small gain would result in a loss in precision and reproducibility. The most common US names that could not be classified for gender were: Wei (1165 papers); Kelly (962); Ali (823); Yu (764); Yi (750); Jun (707); Alex (706); Yang (666); Jing (599); Li (579). For example, Wei is the Romanisation of Chinese male (卫, guard), female (蔷, rose) and unisex (未, future) names and so is unisex in the USA. These examples suggest that the inability of the Western alphabet to reflect the nuances of Chinese names coupled with the recent increase in researchers from China working in the USA (e.g., 4% of US-authored papers in Scopus had an author with the Chinese last name Zhang in 2017 compared to 2% in 2007) has reduced the accuracy of gender classification for recent data.

Since the accuracy of identification of females and males from their first names is different, a random set of 1010 articles with US first authors was drawn using a random number generator and used to estimate the effectiveness of the gender identification (just over 1000 in case of data losses, although these did not occur). This is a novel technique that has not been applied in previous related research, which has analysed uncorrected gender proportions. For each article in the list, the first author gender was identified by a person independent of the project finding their home page and examining their photograph or other text on their home page. Ignoring people for whom this information could not be found (12%), the proportion of the correct number of females and males were calculated and a correction factor calculated. From the 891 authors with a manually identified gender, 558 (63%) portrayed a male gender. Of the 509 human-classified individuals that were also assigned a gender, there were only 16 (3.5%) errors, mostly for Chinese names or names with cultural gender variations (e.g., Valery). Thus, there are few errors when assigning genders but females are more difficult to identify. Nevertheless, the word frequency results are reliable by gender. The people that were not assigned a gender mainly used initials instead of full names or had Chinese or other names associated with cultures that were not widespread in the United States at the time of the 1990 census. Faculty from countries with typically ungendered first names, such as China, as well as countries in which ungendered first names are common, but not the norm, such as India, also account for gender identification problems.

The automatic identification method on the same set of 1010 allocated a gender to 576, with 317(55%) being assigned male. The automatic method underestimates the proportion of male authors and overestimates the proportion of female authors. For statistics related to the proportions of academics by gender, the number of males was multiplied by 1.138 (the actual male percentage divided by the estimated male percentage) and the number of females was multiplied by 0.831 (the actual female percentage divided by the estimated female percentage) to correct for this.

Narrow fields that had fewer than 50 articles in 2017 with a gendered first author were excluded to avoid reporting information about very small fields. The final dataset consisted of 285,619 Scopus journal articles from 285 narrow fields in **2017**, with 157,922 male first authors and 127,697 female first authors (44.7% female; 0.809 females per male).

For RQ2, duplicate copies of articles appearing in multiple subject categories were removed because the tests were across all of academia rather than within individual fields. This left 72018 US male first-authored articles and 57763 US female first-authored articles (44.5% female; 0.802 females per male).

## 4.2  Step 2: Broad and narrow field first author genders

**RQ1**: For each of the 285 narrow fields, the ratio of female-authored to male-authored articles was calculated. Odds ratios (female authored articles/male authored articles) were used to rank fields by degree of gender bias, after correcting for the inaccuracy of the gender identification procedure, as described above.

## 4.3   Step 3: Gendered academic terms

Systematic gender differences across academia were sought by identifying terms that are more likely to be used by one gender and investigating whether they are mainly employed with individual fields or cross-field. Single word terms are more useful to analyse than phrases because the objective was to identify general underlying patterns and phrases are more specific than terms. Topic modelling and co-word mapping were not used because they generate clusters of similar terms in a way that does not allow direct statistical comparisons between different sets because of the need for manual intervention and decisions in the construction of the topics. For example, "testing and validation of a topic model are no sinecures: a topic model needs to be trained on a subset and a number of parameters have to be chosen such as the number of topics to be distinguished (T), the concentration of topics in a document (alpha), and terms in a topic (beta). The results have to be screened manually and/or by automatic scoring systems because topics of poor quality remain possible, also after a number of iterations" (Leydesdorff & Nerghes, 2017).

**RQ2**: Using the dataset with duplicate articles eliminated, the title, abstract and keyword list were segmented into individual words, which were then depluralised (for plural words ending in s only). A word association test consisting of a 2x2 chi-squared statistic (male/female vs. articles with/without term) was calculated to assess the extent to which each of these words associated disproportionately with one gender. The terms with the 1000 largest chi-square values for females were extracted as the base set of gendered terms. The limitation to 1000 was heuristically chosen to avoid having terms that were very rare (and hence overly specialist). The 1000 terms were then ranked in descending order of the ratio of female to male articles and the top 100 investigated qualitatively to fit them to themes (see Step 6). This was repeated for the terms with the 1000 largest chi-square values for males. A Benjamini-Hochberg test (Benjamini & Hochberg, 1995; Holm, 1979) was used on the male and female sets of 1000 terms. This sorts all chi-squared $p$ values in ascending order and uses a formula to pick a threshold, below which all chi-squared null hypotheses are rejected. The test rejected the null hypothesis at alpha=0.001 that any of the terms was not influenced by gender. The smallest of the 1000 chi-squared values with a $p$ value below the threshold was 48.4 for males and 55.7 for females. Thus, it is reasonable to assume that all 2000 terms are gender-related.

## 4.4   Step 4: Field-specific gendered terms

Systematic gender differences within narrow fields were sought by identifying terms that are more likely to be used by one gender in multiple narrow fields. The Step 3 approach could not be applied due to a lack of statistical power.

Word association tests using chi-squared values were calculated as above for each term in the articles of each of the 285 fields to find terms that were more likely to be used by one gender. A separate statistic was calculated for each of the 391,514 terms in the gendered papers, and for each field (i.e., 285 x 391,514 chi-squared values instead of the 391,514 chi-squared values of Step 3), generating a separate list of gendered terms for each field. For example, the term *study* in the narrow field *Optometry* occurs in 28% of male-authored articles and 46% of female-authored articles, giving it a female bias (chi-squared: 4.8). Table 1 illustrates the 20 most female-associated terms for one narrow field. The individual chi-squared values cannot be used for hypothesis tests since this would generate too many (285 x 391,514), each with low statistical power (from using about 1/285 of the data). Instead, terms were identified as important if they occurred in multiple (narrow field) lists. These terms are gendered within fields, for multiple fields.

Table 1. The 20 terms in the narrow field *Community and Home Care* that were most likely to be used by females in comparison to males in article titles, abstracts and keywords as judged by the chi-square statistic. There were 41 articles with a male first author and 124 with a female first author. Percentages are of all same-gender articles.

| Term | Female first-authored articles | Male first-authored articles | Chi-square |
|---|---|---|---|
| nurse | 23 (19%) | 0 (0%) | 8.8 |
| support | 28 (23%) | 3 (7%) | 4.7 |
| home | 26 (21%) | 3 (7%) | 4 |
| need | 30 (24%) | 4 (10%) | 3.9 |
| were | 72 (58%) | 17 (41%) | 3.4 |
| explored | 13 (10%) | 0 (0%) | 3.3 |
| during | 18 (15%) | 1 (2%) | 3.3 |
| palliative | 12 (10%) | 0 (0%) | 3 |
| n | 19 (15%) | 2 (5%) | 3 |
| reserved | 19 (15%) | 2 (5%) | 3 |
| right | 23 (19%) | 3 (7%) | 2.9 |
| experience | 23 (19%) | 3 (7%) | 2.9 |
| end-of-life | 11 (9%) | 0 (0%) | 2.6 |
| all | 29 (23%) | 5 (12%) | 2.4 |
| important | 15 (12%) | 1 (2%) | 2.3 |
| review | 15 (12%) | 1 (2%) | 2.3 |
| end | 15 (12%) | 1 (2%) | 2.3 |
| education | 21 (17%) | 3 (7%) | 2.3 |
| illness | 10 (8%) | 0 (0%) | 2.2 |
| hospice | 10 (8%) | 0 (0%) | 2.2 |

## 4.5   Step 5: Common academic gendered terms

**RQ3**: For each narrow field, the 20 most male associated and 20 most female associated terms were extracted as described in Step 4, using the chi-squared values. The 285 lists of 40 gender-associated words were then combined to give 12,040 terms that each occurred disproportionately often for one gender in at least one field. The choice of the top 20 words rather than the top 10 or 50 was a relatively arbitrary compromise between choosing too few, giving low overlaps, and choosing too many, giving uninformative results. A cut-off chi-squared value was not used because of the different numbers of articles in fields.

Terms selected from multiple fields that usually occurred for the same gender are potential indicators of widespread gender associations in research. For example, the top word *study* was used disproportionately often by females in 108 narrow fields, by males in 5 narrow fields, and was not in the top 20 lists for the remaining 188 narrow fields (see below for more discussion). Thus, females used this term more than males in over a third of academic fields. Individual words can have multiple meanings and usage contexts so the words with the tallies of at least 18 fields were investigated further in Step 6 to identify the reason for their gender imbalances within fields. The figure of 18 was chosen to give an overall total of at least 50 words (giving 56 in total, due to ties). Words with high tallies were only investigated when they occurred at least 70% of the time in the list for the same gender. This is a statistically significant list of terms (see Appendix).

## 4.6   Step 6: Gendered academic themes

**RQ2,3**: The final stage was a subjective qualitative investigation of the word frequency results to fit them into themes. The context of ambiguous terms first needed to be detected because of polysemy and more subtle meaning differences that depend on context. For each term, a random sample of at least 30 matching texts was read either in all fields (RQ2) or in all fields with a relevant gender bias (RQ3), the Key Word In Context (KWIC) method. Word frequency analyses were then used to identify terms that tended to occur in the same documents compared to (i) all other documents, and (ii) other documents by the same gender. The 2x2 chi-squared test was again used to judge significance and rank co-occurring terms. Whereas the (i) test produces a list of words that give insights into the topic of a set of documents by comparing them to all other documents, the (ii) test gives insights into the topic of the documents relative to topics researched by the same gender. This is useful when the (i) test produces a list of generic gendered research topic terms, which the (ii) test factors out.

     To illustrate this, the word *women* (chi-squared 1512 for all fields combined) has the highest female gender bias when co-occurring with *study* and the word *technique* (chi-squared 357) has the highest male gender bias when co-occurring with *study*. In other words, for papers containing *study*, if the term *women* also occurred then the author was probably female and if the term *technique* also occurred then the author was probably male. Another *study*-oriented term for females was *participant*, which occurred in 19.3% of female-authored articles containing *study* compared to 6.4% of female-authored articles not containing *study* for the whole dataset (chi-squared: 5442). Thus, the female-first author context for study includes both *women* and *participant*, suggesting that the apparently neutral term *study* is used by females in the context of researching or studying people.

     After clarifying the meaning of the terms, as above, they were subjectively clustered into themes, separately for RQ2 and RQ3. The clustering used a constant comparison approach, repeatedly comparing the use contexts of apparently similar terms by reading matching titles, abstracts and keywords and conducting word association tests on the terms, comparing the results, to decide whether they should be clustered. All terms were re-compared when clusters were changed to ensure that all clusters were internally consistent after each change. Themes related to people and things were given primacy, when possible. Terms were therefore clustered into alternative themes only when apparently not fitting people or things. This represents one interpretation of the data rather than a definitive characterisation of it. The reason for taking the people/thing dimensions as the default, rather than deriving themes purely from the data (e.g., through independent content analysis coding), was to test its boundaries and to find clear evidence that there were aspects of research that could not be reasonably explained by people/things.

## 4.7   Step 7: Comparison with the people/things theory

The final step, reported in the discussion, is a subjective assessment of whether the results are consistent with the people/things theory. There are many different operationalisations of these dimensions, but the following taken from the US college testing organisation ACT Inc., as previously used in research (e.g., Su, Rounds, & Armstrong, 2009), were chosen for clarity, "**People** (no alternative terms). "People activities" involve interpersonal processes such as helping, informing, serving, persuading, entertaining, motivating, and directing—in general, producing a change in human behavior" (ACT, 1995) and, "**Things** (machines, mechanisms, materials, tools, physical and biological processes). "Things activities" involve nonpersonal processes such as producing, transporting, servicing, and repairing" (ACT, 1995).

# 5   Results

## 5.1   Research question 1: Field participation

There are major differences between broad fields in the rates at which males and females participate (Table 2; for similar author-level results, compare with Figure 1.3 of: Elsevier, 2017). For the broad field of Nursing, there are 1.93 female (US first-authored) articles for every male (US first-authored) article. In contrast, for the Mathematics broad field there are only 0.22 female articles for every male article. There are also substantial variations between the narrow fields within broad fields. Even for the most female-friendly broad field, Nursing, whilst there are nearly 9 female articles for every male Review & Exam Preparation article, males are in the majority for Care Planning articles. The people/thing dichotomy does not fully explain the results (Table 2) because some broad fields, such as Veterinary and Mathematics do not fit into either category. The Veterinary field (not people-related, not thing-related) has high female participation and the l Mathematics field (not thing-related, not people-related) has few females. A complete list of narrow fields is available in the online supplement.

Table 2. The ratio of female to male US first authors for all 26 Scopus broad fields together with the narrow subfields with the highest and lowest ratios (qualification: at least 50 gendered US first authored Scopus journal articles in 2017). F/M odds ratios were multiplied by 0.831/1.138 (the actual male percentage divided by the estimated male percentage / the actual female percentage divided by the estimated female percentage) to correct for gender identification biases. The narrow field list is in the online supplement.

| Broad field | | | Most female narrow subfield | |
| --- | --- | --- | --- | --- |
| | Fields | F/M | Most male narrow subfield | F/M |
| Nursing | 20 | 1.93 | Review & Exam Preparation | 8.68 |
| | | | Care Planning | 0.56 |
| Veterinary | 3 | 1.49 | Small Animals | 2.03 |
| | | | Food Animals | 1.03 |
| Health Professions | 9 | 0.99 | Occupational Therapy | 1.94 |
| | | | Radiological & Ultrasound Technology | 0.49 |
| Psychology | 7 | 0.93 | Developmental & Educational Psychology | 1.50 |
| | | | Psychology (misc) | 0.74 |
| Neuroscience | 9 | 0.82 | Endocrine & Autonomic Systems | 1.24 |
| | | | Sensory Systems | 0.54 |
| Social Sciences | 22 | 0.76 | Gender Studies | 1.86 |
| | | | Political Science & International Relations | 0.31 |
| Immunology and Microbiology | 5 | 0.75 | Virology | 0.86 |
| | | | Applied Microbiology & Biotechnology | 0.54 |
| Medicine | 46 | 0.74 | Obstetrics & Gynecology | 2.02 |
| | | | Orthopedics & Sports Medicine | 0.26 |
| Pharma, Toxicology and Pharmaceutics | 5 | 0.69 | Toxicology | 0.89 |
| | | | Drug Discovery | 0.48 |
| Biochemistry, Genetics and Molecular Biology | 15 | 0.67 | Endocrinology | 1.15 |
| | | | Structural Biology | 0.40 |
| Arts and Humanities | 12 | 0.64 | Museology | 1.08 |
| | | | Philosophy | 0.29 |
| Environmental Science | 12 | 0.52 | Health, Toxicology & Mutagenesis | 0.82 |
| | | | Water Science & Technology | 0.43 |
| Agricultural and Biological Sciences | 11 | 0.49 | Food Science | 0.87 |
| | | | Forestry | 0.37 |
| Chemical Engineering | 7 | 0.47 | Chemical Health & Safety | 0.92 |
| | | | Fluid Flow & Transfer Processes | 0.26 |
| Business, Management and Accounting | 10 | 0.47 | Tourism, Leisure & Hospitality Management | 0.77 |
| | | | Management of Technology & Innovation | 0.35 |
| Chemistry | 7 | 0.39 | Inorganic Chemistry | 0.44 |
| | | | Physical & Theoretical Chemistry | 0.35 |
| Materials Science | 8 | 0.39 | Biomaterials | 0.62 |
| | | | Electronic, Optical & Magnetic Materials | 0.30 |
| Earth and Planetary Sciences | 13 | 0.35 | Oceanography | 0.46 |
| | | | Economic Geology | 0.24 |
| Dentistry | 1 | 0.33 | Oral Surgery | 0.33 |
| | | | Oral Surgery | 0.33 |
| Decision Sciences | 3 | 0.32 | Information Systems & Management | 0.35 |
| | | | Statistics, Probability & Uncertainty | 0.31 |
| Engineering | 16 | 0.32 | Architecture | 0.75 |
| | | | Aerospace Engineering | 0.13 |
| Computer Science | 12 | 0.30 | Human-Computer Interaction | 0.53 |
| | | | Signal Processing | 0.20 |
| Economics, Econometrics & Finance | 3 | 0.28 | Economics, Econometrics & Finance (misc) | 0.39 |
| | | | Finance | 0.23 |
| Energy | 5 | 0.26 | Renew. Energy, Sustainability & the Environment | 0.42 |
| | | | Nuclear Energy & Engineering | 0.18 |
| Physics and Astronomy | 10 | 0.24 | Acoustics & Ultrasonics | 0.36 |
| | | | Physics & Astronomy (misc) | 0.15 |
| Mathematics | 14 | 0.22 | Statistics & Probability | 0.34 |
| | | | Logic | 0.09 |

The Arts and Humanities and Social Sciences are analysed at the level of narrow fields since these seem to be more diverse broad areas than the others in Scopus (Table 3). The thing/people orientation does not explain the Arts and Humanities well. The female-associated area, Museology, ostensibly deals with things – buildings and artefacts – although in practice it also involves communicating, educating people and working with the community (McCall, & Gray, 2014). Whilst most fields could be argued to have a people aspect in the final analysis (e.g., for Aerospace Engineering: aircraft are highly collaboratively built for people to use), individual people that will be met and helped are not always the focus (e.g., Aerospace Engineering is tightly focused on designs and components). Perhaps the only thing-oriented Arts and Humanities narrow field is conservation, although this category also includes some museology journals (e.g., Museum International). Aspects of Music (instruments) and Archeology (artefacts) can focus on things.

Table 3. The ratio of female to male US first authors for Arts and Humanities Scopus narrow fields (qualification: at least 50 gendered US first authored Scopus journal articles in 2017).

| Narrow Field | F/M* |
|---|---|
| Museology | 1.08 |
| Visual Arts and Performing Arts | 1.00 |
| Language and Linguistics | 0.97 |
| Conservation | 0.91 |
| Arts and Humanities (misc) | 0.75 |
| Literature and Literary Theory | 0.72 |
| Music | 0.70 |
| History and Philosophy of Science | 0.58 |
| Archeology (arts and humanities) | 0.46 |
| History | 0.44 |
| Religious Studies | 0.34 |
| Philosophy | 0.28 |

*F/M odds ratios were multiplied by 0.831/1.138 (the actual male percentage divided by the estimated male percentage / the actual female percentage divided by the estimated female percentage) to correct for gender identification biases.

All the Social Sciences are people-related but have vastly different rates of gender participation (Table 4). Like Museology, Library and Information Sciences is ostensibly thing-related field (books, libraries, computers) but in practice focuses on communication and education. It has mostly female first authors. The three partly thing-related fields Archeology (human societies studied through artefacts), Transportation, and Human Factors and Ergonomics (human-related design issues) are male dominated. The power-related themes of management and organisation (although there may be gender differences in the extent to which power is exercised: Appelbaum, Audet, & Miller, 2003) are more common in the male-dominated fields (Sociology and Political Science, Public Administration, Geography, Planning and Development, Transportation, Urban Studies, Political Science and International Relations).

Table 4. The ratio of female to male US first authors for Social Sciences Scopus narrow fields (qualification: at least 50 gendered US first authored Scopus journal articles in 2017).

| Narrow Field | F/M* |
|---|---|
| Gender Studies | 1.86 |
| Health (social science) | 1.45 |
| Life-span and Life-course Studies | 1.33 |
| Education | 1.17 |
| Demography | 1.15 |
| Library and Information Sciences | 1.14 |
| Linguistics and Language | 1.07 |
| Social Sciences (misc) | 1.04 |
| Communication | 0.98 |
| Anthropology | 0.93 |
| Cultural Studies | 0.83 |
| Sociology and Political Science | 0.65 |
| Public Administration | 0.58 |
| Development | 0.58 |
| Geography, Planning and Development | 0.56 |
| Transportation | 0.54 |
| Law | 0.53 |
| Urban Studies | 0.52 |
| Safety Research | 0.48 |
| Archeology | 0.42 |
| Human Factors and Ergonomics | 0.41 |
| Political Science and International Relations | 0.31 |

* F/M odds ratios were multiplied by 0.831/1.138 (the actual male percentage divided by the estimated male percentage / the actual female percentage divided by the estimated female percentage) to correct for gender identification biases.

## 5.2   Research question 2: Overall topics and methods

Irrespective of gendered field choice (RQ1), gendered research topic choices across academia may give a new perspective. The most female-associated terms across academia (Table 5) involve people directly or indirectly. For example, a mailed survey involves contacting people remotely, an indirect connection. Note that, as for the final column of Table 2, this section focuses on extreme cases of gender imbalance and should not be inferred to cover all types of research.

Table 5: The 100 terms most likely to be used by females in comparison to males for the overall Scopus 2017 dataset (after duplicate article elimination) organised by subjectively determined theme. F:M is the proportion of female-authored articles containing the term divided by the proportion of male-authored articles containing the term.

| Theme | F:M* | Statistically significant gendered top 100 terms** |
|---|---|---|
| Mothers | 15.02 | Mothering, motherhood, mother-child, mother, maternal |
| Babies | 13.21 | Breastfeed, breastfeeding, birthweight, babies |
| Children | 9.79 | Preschool-aged, pre-schooler, childcare, toddler, parenthood, teen, parenting, maltreatment, pubertal |
| Childbirth | 8.65 | Pre-pregnancy, childbirth, GDM, postpartum, antenatal, trimester, Cesarean, perinatal, pregnancy, NICU, gestational pregnancies, pregnant, maternity |
| Women | 7.74 | Latina's, herself, feminist, feminism, Latina, femininity, |
| Interpersonal | 6.56 | Mother-infant, family-centered, parent-child, interprofessional |
| Talking | 6.38 | Talked, speech-language, audio-recorded, verbatim, linguistically |
| Education | 5.67 | Baccalaureate, service-learning, practicum, preschool, mentoring |
| Psychiatric illness | 5.16 | Bulimia, nervosa, eating, trauma-exposed, trauma-informed, telehealth, internalizing, externalizing |
| Gender inequality | 4.25 | Sexism, gendered |
| Sexual violence | 3.94 | IPV, rape, harassment |
| Contraception | 3.67 | Contraception, contraceptive |
| Female health | 3.53 | Menopausal, menopause, menstrual, gynecology |
| Interviews | 3.34 | Videotaped, facilitator, semistructured, transcribed [see: Talking] |
| Nursing | 3.26 | Nursing, nurse |
| Psychotherapy | 3.24 | Yoga |
| Social inequality | 3.06 | Socioeconomically, intersectionality, sociocultural, underserved |
| Food health | 3.04 | Lunch (e.g., national school lunch programmes) |
| Health management | 2.91 | Self-management |
| Language community | 2.78 | Monolingual, Spanish-speaking |
| Carers | 2.73 | Caregiving, caregiver |
| Survey | 2.54 | Mailed |

*Multiply by 0.809 to get the F/M author ratio.

**Ambiguous (e.g., polysemous) terms occurring within multiple possible themes are not shown: Mulatta, self-perceived, macaca, culturally, cooking, coaching, work-related (e.g., 'Mulatta' occurred in different uses of Macaca mulatta rhesus macaques in experimental and wild contexts).

Some of the most female-associated terms reflect perspectives that are already evident from field choices. The Mothers, Childbirth and Babies themes reflect the female dominance of Maternity and Midwifery (female-authored articles are up to 6.33 more common than male-authored articles in this narrow field), a narrow field within Nursing, but extended to other fields. The Children theme also partly reflects Pediatrics (6.62 F/M) within Nursing. The Education theme (5.67 F:M, which translates to 5.67×0.809= 4.59 F/M) gives a much stronger gender imbalance than for the Education narrow field (1.17 F/M, i.e., 1.17 female first-authored articles for every male first-authored article). Two psychology-related themes are Psychiatric illness and Psychotherapy, where the female dominance contrasts with the much lower female dominance in Clinical Psychology (1.05 F/M). Gender inequality also echoes Gender Studies (1.86 F/M).

In terms of new perspectives provided by the top terms list, the Women theme echoes female dominance of Gender Studies (1.86 F/M) within the Social Sciences but provides a different angle. Sexual violence is not an academic field, but relates to Gender Studies, Sociology, Law and Health (Social Science). Contraception relates to Reproductive Medicine (1.31 F/M; excluded from main analyses for having only 46 qualifying articles). Female health aligns with Obstetrics and Gynecology (2.02 F/M; excluded from main analyses for having only 46 qualifying articles). Carers contrasts with Community and Home Care (0.55 F/M; excluded from main analyses for having only 20 qualifying articles). Social inequality, Food health, and Survey also give new dimensions.

The most male-associated terms (Table 6) are much more specialist than the female-associated terms and more closely follow broad or narrow fields. For example, the term homotopy (a map linking two continuous functions between two topological spaces) might be explained in the third year of a mathematics degree and is specific to pure maths.

Table 6: The top 100 male-associated academic terms for the overall Scopus 2017 dataset organised by subjectively determined theme. M:F is the proportion of female-authored articles containing the term divided by the proportion of male-authored articles containing the term.

| Theme | M:F* | Statistically significant gendered top 100 terms (up to 4)** |
|---|---|---|
| Pure maths | 40/0 | Countable, Riemannian, homotopy, axiom... |
| Engine component | 15.74 | Coolant, rotor, actuator, rotator,... |
| Physics | 15.59 | Relativity, LHC, astrophysical, spacetime,... |
| Measurement | 13.40 | Mach |
| Surgery | 12.08 | Periprosthetic, arthroscopic, decompression, embolization |
| Bone surgery | 9.72 | Arthrodesis, TKA, acetabular, humeral,... |
| Signal processing | 7.45 | Interferometer, resonator, khz, mhz, ... |
| Scholarly debate | 7.32 | Reply, erroneous |
| Computing | 6.53 | Arbitrarily, open-source |
| Medical imaging | 6.45 | Angiographic |
| Economics | 5.64 | Macroeconomic, liquidity, inflation |
| Engineering | 5.04 | Vorticity |
| Religion | 4.20 | Doctrine |

*Divide by 0.809 to get the F/M author ratio

**Ambiguous (e.g., polysemous) terms occurring within multiple possible themes are not shown: PCI, loosening, superposition, rectangular, instabilities, vortex, convective, rotational, mesoscale, drag (e.g., PCI could mean percutaneous coronary intervention or prophylactic cranial irradiation).

Many of the terms and themes are not deducible from narrow fields, including talking, women, interviews, children and engine components. This shows that the RQ2 perspective adds to RQ1.

## 5.3 Research question 3: Within field general topics and methods

Terms that are gendered within narrow fields bypass field choices to some extent and may give insights into extreme cases of gendered interests or choices made after selecting a narrow field to investigate, although the results are also influenced by multidisciplinarity. The words that were gendered in at least 17 narrow fields and with at least 70% being female-associated (Appendix: Table A11) were grouped subjectively into themes after checking the contexts in which the terms occurred within titles, abstracts of keywords from the data sets (Table 7). Papers from 2017 with US female first authors are more likely to employ each term than papers from 2017 with US male first authors from the same narrow field in at least 17 fields. Some of the listed themes overlap (e.g., qualitative methods and people-related methods). The gendered terms for each theme are terms in Table A11. Terms in square brackets are indirect (i.e., not directly expressing the theme but nevertheless used when discussing the theme, such as *support* in the context of supporting people) or shared with other contexts (e.g., *gender* is often used in research focusing on females but is also used when discussing LGBTQ issues).

Table 7: Themes detected by using word association tests to explore the contexts in which *female* gendered terms in Table A11 were used in titles, abstracts or keywords. Terms in square brackets are often mentioned when the theme is mentioned but do not directly describe the theme, or sometimes describe unrelated concepts.

| Theme | Terms gendered in at least 17 narrow fields |
|---|---|
| Qualitative methods | study, interview, qualitative, [finding, result, reported] |
| Exploratory methods | explore, examined, examine, [among, with] |
| People | women, children, gender, mother, parent, social, community, age, behavior, practice, [support, experience] |
| > Social groups | social, community |
| > Females | women, mother, [gender] |
| > Children | children [mother] |
| > Parenting | parent [mother, children] |
| Education | education, student, [practice] |
| People-related methods | interview, participant, intervention, experience, group, [support] |
| Cell biology | cell |
| Environmental and other impacts | impact |
| Hedging argument style | may |
| Care | care |

A few themes (Table 8) were identified from the contexts of the male-associated terms in Appendix Table A12. There is not a strong thing orientation in the results but this might be because things differ between fields and do not have many generic terms that are likely to be used in academic papers. Male cross-topic themes are less widespread than female cross-topic themes, possibly because there are many more different words for "things" in fields than for people, or because quantitative research has more specific and varied terminology.

Table 8: Themes detected by using word association tests to explore the contests in which *male* gendered terms in Table A12 were used in titles, abstracts or keywords. Terms in square brackets are often mentioned when the theme is mentioned but do not directly describe the theme, or sometimes describe unrelated concepts.

| Theme | Terms gendered in at least 17 narrow fields |
|---|---|
| Quantitative methods | simulation, performance, measurement, value, small, [show]. |
| Things | it |
| Abstract analytical approach or method | [then], model, simulation, [paper]. |
| Experiments | experiment, simulation |
| Patients being treated | patients |

## 6   Discussion

Some methodological limitations have already been noted, including cultural bias from the first name algorithm, occasional inappropriate Scopus subject classifications, and using publications as a proxy for participation in research. Moreover, the analysis largely ignores external factors other than topic that may shape career decisions, such as chilly climates in some fields. The gender correction method (using online web profiles to check author gender for a random sample) assumes that the first name gender detection method is equally biased in all fields. It is possible that some fields have (a) a relatively high (or low) proportion with names that were not detected for one gender, for example if the field had attracted a high proportion of overseas researchers (whose first names would be less likely to be resolved for gender) and (b) these international researchers had a different gender balance than the US-born researchers within the field. Another important limitation is that the dataset combines senior and junior researchers, obscuring changes that have occurred over the past 40 years. Results for current PhD students, for example, may show smaller, or different, gender differences. The results also do not address gendered culture differences within fields that do not translate into print, or that manifest in non-Scopus (e.g., regional or low citation impact) journals. The results may be affected by the inclusion of internationally collaborative studies with a US first author because gender roles differ internationally and may be affected by gender differently in mixed gender teams or on fields, such as medicine, in which the last author may determine the research topic. The term-based results (RQ2, RQ3) are a limitation because an individual word can have different meanings within different phrases (e.g., "Shakespeare play" and "play time"). This was guarded against by the word association analyses to interpret the terms but reduces the statistical power of the tests used for RQ2 and RQ3. The word frequency results are affected by the specificity of terms used in fields. Topics that use specialist terminology more, such as molecule names, are less likely to produce significant terms in the RQ2 and RQ3 analyses than topics that avoid such terms or use more general terms in parallel. Thus, the topic terms found should not be regarded as a complete list or even the most important topics. Instead they represent a non-random sample of topics with statistically significant evidence of gender difference. Finally, whilst the discussion focuses on gender imbalances and extreme cases, it should not be forgotten that some broad fields have approximate gender parity.

Research publishing reflects many factors besides interests, such as the availability of journals and editorial policy decisions. In addition, a researcher may write papers in areas that they believe are publishable, which do not necessarily reflect their interests. There may even be systematic factors that result in academics researching vocational areas that they dislike, choosing research as a way of avoiding an occupation that they have trained for. For example, some nurses might join the profession due to family pressure or job availability, then leave practice to research after finding that they dislike interacting with sick people. In the main analyses of the current research, such people would be incorrectly assumed to want a people-oriented career.

The results for RQ1 and RQ3 are also limited by the Scopus classification scheme used. Thus, although the concept of a narrow field seems intuitively meaningful, in practice fields overlap and incorporate varying degrees of interdisciplinarity. Journals also incorporate varying degrees of interdisciplinarity. For example, an Optics paper about nursing might have been published in *Journal of Ophthalmic Nursing & Technology* and could therefore be considered as a cross-field paper. The choice of year may also affect the results given that gender differences change over time. The core word frequency approach may obscure gender differences or exaggerate the importance of others. Field specialty topics do not necessarily indicate the reason why researchers have chosen them: One production engineer might be interested in the mechanics of production processes whereas others might be fascinated by the social organisation involved or the availability of jobs in their area.

Table 9 suggests some themes through which the results can be interpreted. Given the multifaceted nature of language and concepts, they are not definitive but are (a) consistent with the data and (b) informed by the people/thing theory.

Table 9: Female-associated themes and examples of supporting evidence from the three dimensions.

| Theme | Broad or narrow field (Tables 2,3,4) | Overall research topic terms (Table 5) | Within-field topic terms (Table 7) |
|---|---|---|---|
| People | | | |
| >Females | | Mothers (F:M up to 15.02); female health (F:M up to 3.53) | *women* |
| >Babies/maternity | Maternity & Midwifery (F/M: 6.33) | Baby-related (F:M up to 13.21); childbirth (F:M up to 8.65) | |
| >Children | | Children (F:M up to 9.79) | *children* |
| >Groups | Demography (F/M: 1.15) | Lang. comm. (F:M up to 2.87); social inequal. (F:M up to 3.06) | *community, social* |
| Caring/nurturing | | | |
| >Human care | Nursing (F/M: 1.93) | Psychotherapy (F:M up to 3.24); health man. (F:M up to 2.91) | *care* |
| >Education | Education (F/M: 1.17) | Education (F:M up to 5.67) | *education* |
| >libraries | Library & Info. Sci. (F/M: 1.14) | *librarian* (F:M: 2.6) | |
| >Applied psychology | Developmental & Edu. Psych. (F/M: 1.50) | Psychological illness (F:M up to 5.16) | |
| >Human communication | Linguistics and Language (F/M: 1.07) | Talk (F:M up to 6.38); interpersonal comms. (F:M up to 6.56) | |
| >Environmental & other impacts | | | *impact* |
| Qualitative & exploratory meth. | | | *qualitative, explore* |
| Interviews & surveys | | Surveys (F:M up to 2.45); interviews (F:M up to 3.34) | *interview* |
| Veterinary science | Veterinary (F/M: 1.49) | | |
| Hedging | | | *may* |
| Cell biology | Cell Biol. (F/M: 1.04) | | *cell* |

## 6.1 Research question 4: Female topics - are people enough?

The people dimension is defined here as involving, or directly related to, individuals or groups of humans. All three analysis perspectives (field choice, overall terms, within field terms), give results for females that are consistent with the people/thing theory hypothesis that females are more likely to research people-related topics. The evidence always points to types of people or activities involving people rather than people in general (e.g., *person,* *people* were not top female topic terms), but this may be a method limitation rather than a general principle. The evidence suggests the **refinement** that some types of people attract a stronger female gender bias, although this could be due to either less research overall, a smaller gender bias, or both. Thus, it is possible, but not proven by the data, that the people concept is too general and may need refinement. There is strong evidence that females research aspects of females, children, babies and groups more, although not necessarily all aspects in each case, as well as human communication.

Caring and nurturing is another **refinement** on the people concept. There is substantial evidence of a variety of female-associated activities that could be characterised as caring or nurturing, including care, education, libraries, applied psychology and (human) communication. A concern with impact on people also loosely fits into this category.

Qualitative and exploratory methods as well as interviews and surveys both involve people and are part of research projects that are about, or related to, people. Their inclusion could therefore be a **side-effect** of a general female people-focus. Alternatively, they may be an attractive low-cost alternative for females given that senior (mainly male) academics dominate funding in some areas (Levitt & Levitt, 2017).

The veterinary theme is not human-related and is an **exception** to the people/thing theory. Veterinary science involves caring for, or treating, non-human animals. The four terms most strongly associating with the Veterinary broad category are all equine (mare, horse, equine, stallion), and so the female interest may be primarily driven by horses. Females increased from 10% of veterinary medical college students in 1971 to 80.5% in 2017. This is largely not matched at the faculty level, with 36% female tenure track faculty, 59% female non-tenure track clinical and 50% female non-tenure track research faculty (AVMC, 2017). Females have been found to have a more positive attitude towards animals than males, with slightly stronger attachment to pets and substantially greater opposition to cruelty and hunting (Herzog, 2007). Whilst veterinary science does not focus on interactions with people, it presumably shares the trait of kindness that is associated with people-oriented job interests (see Table 9.1 of: Spokane, Luchetta, & Richwine, 2002). According to the US Department of Labor sponsored O*NET (onetonline.org), Veterinarians (code 29-1131.00) and Animal Scientists (19-1011.00) have Realistic and Investigative job types, which would score as a thing-oriented job.

Cell biology is a second **exception** to the people/thing theory. The gender difference is of marginal practical significance in the Cell Biology (F/M: 1.04) narrow field, but females were a larger majority in the related narrow fields of Developmental Biology (F/M: 1.54) and Endocrinology (F/M: 1.57), and the cell biology theme has a common within-field female bias. Thus, there seems to be a consistent female interest in cell biology in comparison to, say, the whole organism topics of Insect Science (F/M: 0.44) or Aquatic Science (F/M: 0.65). This may relate to human biology interests. Female-authored Cell Biology papers were slightly more likely to mention cells, gene expressions, fertility, progesterone (a sex hormone for humans and some other animals) and profibrotic. Cell Biology may be perceived as being likely to provide the communal goal of social impact (through curing diseases), which may attract more females (Diekman et al., 2017). For example, females interested in a STEM career may be more likely to choose cell biology and other life sciences that seem to have the potential to deliver social impact. There may also be a second-order effect due the increase in the proportion of females studying biology and the increased focus on cells and genetics. Thus, younger researchers, who are predominantly female, may be more likely to research cell biology rather than other areas of biology. Overall, however, the data gives little insight into the female association of cell biology. O*NET classifies Molecular and Cellular Biologists (19-1029.02) with Holland RIASEC codes Investigative, Realistic and Artistic (ignoring Social, Enterprising, Conventional), suggesting that this is a thing-oriented job.

## 6.2   Research question 4: Male topics - are things enough?

The thing dimension is defined here as involving, or directly related to, natural or human-made physical objects. The object could be part of a larger whole (e.g., engine) and could be conceptual, such as a computer system. A greater male tendency to research things could explain some, but not all, of the results (Table 10). The variety of possible terms for things may be why the thing concept cannot be refined to the extent that the people concept could be for females. Computing and Medical imaging could be classed as about things, although the male-oriented computing term *open-source* is not a thing but sits inside a computer.

Table 10. Male-associated themes and examples of supporting evidence from the three dimensions.

| Theme | Field choice (Table 2,3,4) | Overall research topic terms (Table 6) | Within-field topic terms (Table 8) |
|---|---|---|---|
| Things | Physics & Ast. (F/M: 0.24); Materials Sci. (F/M: 0.39); Engineering (F/M: 5.04) | Engine components (M:F up to 15.74) | *it* |
| >Computing | Computer Sci. (F/M: 0.30) | *open-source* (M:F 7.32) | |
| >Medical imaging | Radiology, Nuclear Med. & Imaging (F/M: 0.39) | *angiographic* (M:F 6.45) | |
| Patients and surgery | Surgery (F/M: 0.38) | Surgery (M:F up to 12.08) | *patients* |
| Power/control | | | |
| >Politics | Political Sci. & International Rel. (F/M: 0.31) | | |
| >Business | Business, Accounting & Man. (F/M: 0.47). | | |
| >Planning | Geog., Plan. & Dev. (F/M: 0.56) Urban Studies (F/M: 0.52) | | |
| >Safety | Safety Res. (F/M: 0.48) | | |
| >Law | Law (F/M: 0.53) | | |
| >History | History (F/M: 0.44) | | |
| >Economics | Economics, E. & Fin. (F/M: 0.28) | Economics (M:F up to 7.54) | |
| >Religion | Religious St. (F/M: 0.34) | *doctrine* (M:F 4.2) | |
| Quantitative methods | Math. (F/M: 0.22) Decision Sci. (F/M: 0.32). | *mach* (M:F 13.40); signal processing (M:F up to 7.54) | *measurement* |
| Experiments | | | *experiment* |
| Abstract analytical approaches | Phil. (F/M: 0.28), Math. (F/M: 0.22) | Pure maths (M:F up to 40/0) | *model* |
| Scholarly debate | | *reply*, *erroneous* (M:F up to 7.32) | |

Males using quantitative methods mirrors females using qualitative methods and is possibly a **side-effect** of a greater male focus on things, presumably often measuring them. The same may be true for experiments. Alternatively, it may stem from childhood stereotypes breeding greater mathematical confidence among males.

Patients and surgery are an **exception** in terms of a people-related themes that are researched more by men (this could also be classed as one theme). It is possible that patients are sometimes objectified in the context of medical treatments, with the disease being the primary research focus rather than the sufferer. Nevertheless, research involving patients or surgery directly relates to people even if could be conceived as being not primarily focused on them. Moreover, research about patients or surgery is also caring in the sense that its ultimate objective is to directly help people. It is possible that surgery is perceived as being more likely to provide the agentic goal of high status, which may attract more males (Diekman et al., 2017). It is not clear whether this would apply to surgery more than to other areas of medicine, however, although some areas of surgery are highly paid[2].

A second **exception** to the people/things theory is that several other of the male-associated research fields ostensibly involve people but do not have an obvious thing aspect. Some of these *could* be generalised as research topics with a power or controlling dimension. This is clearest for politics, but is also an aspect of business, planning, safety research (e.g., safety regulations), and law. For example, Industrial Relations (F/M: 0.42) is a people-related narrow business field. Some history research is about politics or situations involving the exercise of power (History articles contain the terms *political* [21%], *war* [13%] and *empire* [5%]), and so it could also fit into this category. Research into patients and surgery may also have a power/control element since medical professionals have power over the lives of their patients. Power is more important to male career choices (Gino, Wilmuth, & Brooks, 2015). Again, expertise in these fields may be perceived as being more likely to satisfy a high-status goal, attracting more males (Diekman et al., 2017).

There is greater male interest in religious research, which can have a controlling dimension, although religion can also have caring aspects and religious debates may be abstract (see below). Women form 17% of the clergy in the USA (BLS, 2018). Although they are a higher proportion of wider religious workers (e.g., 65% of "Religious workers, all other" in the same survey) presumably these are much less likely to need religion-specific higher education.  The dominance of male researchers for religion may therefore be, in part, a side effect of education for religious vocations being primarily for men, rather than any controlling or abstract dimension of religion. This would give a greater pool of males with a religious education from which to draw qualified educators and researchers. Status is again a potential explanation (Diekman et al., 2017).

Another people/things **exception** is that the abstract fields of Mathematics and Philosophy both have a high proportion of male first authors, despite an absence of things. Most research mathematics has a quantitative element but pure mathematics is characterised by extreme abstraction (exemplified by the terms *axiom* and *homotopy*) rather than numbers. Taken together, they suggest a male abstraction dimension. In support, physics (within Physics and Astronomy) also deals with abstractions (e.g., relativity), albeit with quantitative elements, and is strongly male. The male associated term *model* also signifies abstraction and seemed to be typically used in the context of theoretical (rather than physical) models. The term *theory* is a clearer indicator of this, although it fell below the threshold for inclusion (only 15 fields). *Theory* occurs substantially more for male articles in diverse areas but mainly in Psychology (3 fields; also 2 Neuroscience). A male interest in abstraction does not seem to have been noted before. Males discuss metaphysical concepts (e.g., God) more than females (Newman et al., 2008) but there is a slight female preference for ideas rather than data (Su, Rounds, & Armstrong, 2009). Ideas are a type of abstraction but this only shows that females prefer abstraction (ideas) to data. There does not seem to be a gender difference in the ability to understand abstract ideas (Park, Hyun, & Heuilan, 2015; Zeitoun, 1989). A possible explanation is that abstract fields seem intuitively less likely to be perceived as supporting the communal goal of social impact. If true, this would help to explain fewer active females (Diekman et al., 2017; Yang & Barth, 2015). These results partly contrast with Su and Rounds' (2015) meta-analysis finding of a relatively small gender difference effect size in RIASEC interests of d=0.34 (males more interested) compared to d=0.36 for science and d=1.11 for engineering. It therefore seems likely that abstract mathematics research careers are much less attractive to females than mathematics as a subject, perhaps because of the ready availability of mathematics teaching as an alternative career, satisfying the social impact communal goal (Diekman et al., 2017) and delivering a people-oriented career.

The male preference for scholarly debate (e.g., reply, erroneous) as a possible exception does not seem to have been noted before but relates to a small number of articles (160 altogether), and may be a **side-effect** of male-associated interests in law, philosophy and religion.

# 7    Conclusions

There are substantial gender differences in US academic first authoring of journal articles across academia between broad fields and between narrow fields within broad fields (RQ1). Many terms were gendered across all fields of science (RQ2) or were gendered within fields of science (RQ3). The combined results of the three different methods have suggested refinements, exceptions and side-effects to the people/thing theory but not the causes of research choices (e.g., biology-influenced preferences, sexist constraints, education, societal gender pressures, work-life balance constraints, field cultures, differential access to resources). The results of this paper are a subjective interpretation of the quantitative data, attempting to give plausible new perspectives on gender differences in academia and assessing the extent to which the people/thing theory is adequate to explain research publishing in the USA. All the findings in this paper are about the extent of gender differences rather than binary differences. There was only one minor example of a monogender field or theme (a type of pure maths in Table 6).

The previously hypothesised people/thing dimensions are broadly consistent with *some* of the findings, in the sense they *could* explain some of the gender differences found. The female-people association may not apply equally to all people, but is strong for the study of maternity, females and children, and for nurturing/caring. It is not universal because it *reverses* for control/power-related topics. It does not explain the female association with veterinary or cell biology research. The thing dimension matches some male preferences but not the male association with **abstraction** and **power/control** topics. Thus, the people/things dimensions can only provide a partial explanation for gender differences in topic choices across the full spectrum of academia because there are many important exceptions. Given the power/control exception to the people/thing dimensions, **Helping/caring/nurturing** could be more precise than people as a dimension. A different research approach would be needed to test this. Whilst individual exceptions to the people/thing theory could be explained by historical factors, the trends in the exceptions suggest that they might be more fundamental. It is possible that most exceptions here to the people/thing theory could be explained by gender differences in agentic (status) and communal (social impact, family) goals (Diekman et al., 2017), as discussed above, although evidence is needed to test whether combining these two theories is enough. The prior hypothesis that disciplinary cultures in some quantitative fields (Cheryan, Ziegler, Montoya, & Jiang, 2017) and surgery (Yu, Jain, Chakraborty, Wilson, & Hill, 2012) can alienate females is insufficient to account for some of the deviances found here, such as the male dominance of some people-related fields. Given that the current research has not attempted to assess any cause and effect relationships, deviations from the people/thing dimensions could also be due to other factors within academia that deflect people from pursuing their interests, such as editorial, departmental or funding policies.

The results also suggest that the things and people dimensions (and other gender dimensions, such as abstraction) may occur within fields as well as in field choice. This can occur when female researchers tackle the human dimensions of male-associated fields, such as in Electrical and Electronic Engineering (e.g., 'The Underrepresentation of Women in Computing Fields: A Synthesis of Literature Using a Life Course Perspective'), Development (e.g., 'Experiences of adolescent mothers in Costa Rica and the role of parental support'), and Economics and Econometrics (e.g., 'Fundraising as women's work? Examining the profession with a gender lens'). It can also occur when males take a more abstract approach (e.g., with a theoretical model or simulation) than typical for a field.

To the extent that the gender differences found are due to the genuinely free choices of the researchers and not limited by social constraints (Webster, Rice, Christian, Seemann, Baxter, Moulton, & Cil, 2016), they are not problematic from a gender equity perspective. They may still be a concern for academic vitality if areas of scholarship lack input from one gender.

Although this article has investigated gendered research patterns rather than their causes, the results suggest some possible new remedies for gender imbalances. With this assumption, a practical application of the findings is that research managers and journal editors in fields for which women are underrepresented may consider, when relevant, embracing multidisciplinary research with areas that attract more women, and may consider being more open to qualitative methods and people-centred topics. For example, a quantitative product development engineering journal might benefit from interview-based research about how key techniques are used in practice (e.g., Rossi, 2017). The apparently greater female use of exploratory methods (perhaps from small group research and interviews) suggests that scientific policies giving primacy to more positivist paradigms may disproportionately disenfranchise females. Whilst positivism is essential in some fields (e.g., physics) it can be a debating point in others (e.g., human-computer interaction). Research group leaders and editors may also reflect on whether they are over-reliant on abstraction or at risk of a research-practice gap, both of which may disadvantage females. Employers may also consider making similar decisions to attract and keep female researchers when they are underrepresented. These recommendations are simplifications that need to be applied sensitively to individual subjects and contexts, especially given the lack of cause-and-effect relationships for the gendered patterns found here. They are additional to known best practice strategies for hiring and retaining females that consider likely caring responsibilities, the right to switch between full-time and part-time working, extra support after career gaps, ignoring career gaps due to childbirth or carer responsibilities when considering tenure or promotion, and supporting males adopting carer roles. They should also complement attempts to project more broadly inclusive and female-friendly cultures and ensure that girls have early and positive exposure to traditionally masculine fields.

# 9 Appendix A: Extra tables

Table A11. Female-associated words occurring in the top 20 male or female words of at least 17 different narrow fields (70%+ female) for USA first-authored Scopus articles in 2017. Context was deduced from co-occurrences: examining words that frequently occur in the same documents as the given term for females in comparison to either (a) the same term for males, or (b) different terms for females.

| Term | Fields | Female | Context in which the term is often used in the dataset |
|------|--------|--------|--------------------------------------------------------|
| study | 73 | 100% | Mixed - Method (e.g., qualitative, case study; multiple) |
| women | 56 | 98% | People-related topic, female-associated topic |
| were | 51 | 94% | Mixed - method (e.g., involving people; "[] were…") |
| participant | 46 | 98% | People-related methods |
| children | 41 | 100% | People-related topic |
| interview | 41 | 100% | People-related method |
| health | 41 | 95% | Caring, indirectly people-related |
| was | 35 | 91% | Mixed - method (e.g., involving people; "[] was…") |
| intervention | 31 | 100% | people -oriented method |
| experience | 30 | 100% | Mixed - people-related discussion |
| their | 30 | 87% | Topic, people-related |
| education | 29 | 100% | People-related topic |
| qualitative | 27 | 100% | Method, associated with interviews, indirectly people-related |
| research | 27 | 96% | Method description style associated with qualitative research, indirectly people-related |
| examined | 26 | 96% | Examining associations - exploratory methods |
| child | 25 | 100% | People-related topic |
| student | 25 | 100% | People-related topic |
| group | 24 | 88% | People-related method (participants); randomized control group method |
| care | 23 | 91% | Caring, indirectly people-related |
| among | 22 | 91% | Examining associations among variables or factors - exploratory methods |
| finding | 21 | 100% | Method description style associated with qualitative research, indirectly people-related |
| result | 21 | 71% | Method description style associated with qualitative research, indirectly people-related [quantitative articles were less likely to use this term in abstracts, titles and keywords, despite often having results] |
| gender | 20 | 100% | People-related topic |
| explore | 20 | 95% | Exploratory methods |
| examine | 20 | 90% | Examining associations among variables or factors (i.e., exploratory methods) |
| with | 20 | 75% | Examining associations among variables or factors (i.e., exploratory methods) |
| support | 20 | 70% | Method and issue - support in interview [verb], social support [noun] |
| mother | 19 | 100% | People-related topic |
| parent | 19 | 100% | People-related topic |
| cell | 18 | 100% | Life-oriented topic |
| community | 18 | 94% | People-related topic |
| social | 18 | 94% | People-related topic |
| behavior | 18 | 89% | People-related topic - mainly human behaviour |
| impact | 18 | 78% | Outcome, particularly climate change and environment |
| age | 17 | 94% | People-related topic aspect |
| practice | 17 | 94% | Educational and professional practice, particularly relating to care |
| reported | 17 | 88% | Methods – qualitative (participants reported) and surveys |
| may | 17 | 76% | Hedging conclusions |

Table A12. male-associated words occurring in the top 20 male or female words of at least 17 different narrow fields (70%+male) for USA first-authored Scopus articles in 2017. Context was deduced from co-occurrences: examining words that more frequently occur in the same documents as the given term for males in comparison to either (a) the same term for females, or (b) different terms for males.

| Term | Fields | Female | Context in which the term is often used in the dataset |
|---|---|---|---|
| is | 28 | 11% | Mixed - quantitative orientation |
| model | 27 | 11% | Thing (abstract) |
| it | 24 | 4% | Thing (anaphor) |
| patient | 23 | 26% | Caring, indirectly people-related, mainly related to surgery |
| value | 23 | 4% | Method, quantitative |
| an | 22 | 23% | In the context of a thing |
| measurement | 22 | 5% | Method, quantitative |
| simulation | 21 | 29% | Thing (abstract) |
| then | 20 | 0% | Abstraction orientation (if, then hypotheticals) |
| experiment | 19 | 21% | Method - experiment with things or simulation |
| show | 18 | 28% | Quantitative and abstract (Results or simulation shows) |
| small | 18 | 28% | Method (Size) and topic (molecule) |
| these | 18 | 22% | Unclear |
| paper | 18 | 6% | Abstract writing style associating with abstract methods (e.g., "this paper…") |
| performance | 17 | 29% | Quantitative orientation: performance of system or model |
| some | 17 | 6% | Quantitative orientation (e.g., "some of the channels") |

# 10 Appendix B: Step 5 statistical significance check

A cross-check was conducted to test whether any of the selected words were likely to have occurred at random. The minimum overall frequency of any word in the selected set was 2783 (the word *nursing* occurred in 2783 different articles) and 2613 terms occurred in at least 2783 articles. Ignoring lower frequency words (a conservative approach), and assuming that the frequency distribution of each word is independent between fields (which is not true since many journals are in multiple fields) then the probability that the same word occurs in the top 20 lists of at least 17 of the 285 fields can be calculated exactly using the binomial distribution formula Binomial(n=285, p=20/2613). The probability that any of the 2783 terms occurs in at least 17 fields is 0 to 6 decimal places. Journal overlap could be modelled conservatively by assuming that each journal appears in 2.2 different categories (the average for the current data) and repeating the calculations but dividing the number of fields required by 2.2: any word occurring in at least 17/2.2=8 (approx.) fields out of 285/2.2=130. The probability of this is 0.026. This it is likely that none of the 56 selected gendered terms has a spurious cause.